\documentclass[prd,twocolumn,superscriptaddress]{revtex4}
\usepackage{multirow,graphicx,verbatim}
\usepackage{subfigure}

\newcommand{\met}{\mbox{${\hbox{$p$\kern-0.4em\lower-.1ex\hbox{/}}}_T \!$ }}
\newcommand{\metx}{\mbox{${\hbox{$p$\kern-0.4em\lower-.1ex\hbox{/}}}_x \!$ }}
\newcommand{\mety}{\mbox{${\hbox{$p$\kern-0.4em\lower-.1ex\hbox{/}}}_y \!$ }}
\newcommand{\vecmet}{\mbox{${\hbox{$\vec{p}$\kern-0.4em\lower-.1ex\hbox{/}}}_T \!$ }}

\begin{document}

\title {Higgs boson coupling sensitivity at the LHC using $H\rightarrow\tau\tau$ decays}
\affiliation{University of Oxford, Oxford OX1 3RH, United Kingdom}
\affiliation{University of Warwick, Coventry CV4 7AL, United Kingdom}
\author{Christopher Boddy}
\affiliation{University of Oxford, Oxford OX1 3RH, United Kingdom}
\author{Sinead Farrington}
\affiliation{University of Warwick, Coventry CV4 7AL, United Kingdom}
\author{Christopher Hays}
\affiliation{University of Oxford, Oxford OX1 3RH, United Kingdom}

\begin{abstract}
We investigate the potential for measuring the relative couplings of a low-mass Higgs boson at the 
Large Hadron Collider using $WH$, $ZH$, and $t\bar{t}H$ production, where the Higgs boson decays to 
tau-lepton pairs.  With 100 fb$^{-1}$ of $\sqrt{s} = 14$ TeV $pp$ collision data we find that these 
modes can improve sensitivity to coupling-ratio measurements of a Higgs boson with a mass of about 
125 GeV/$c^2$.
\end{abstract}
\maketitle

\section{Introduction}
The recent discovery \cite{discovery} of a resonance with a mass of about 125 GeV \cite{cdef} in $pp$
collisions at the Large Hadron Collider (LHC) could well correspond to the long-awaited observation
of the Higgs boson \cite{Higgs} of the standard model (SM).  If so, it would herald another remarkable
success of the SM, which predicted the existence of a Higgs boson with a mass less than 152 GeV
at 95\% confidence level (C.L.) \cite{lepewwg} based on precision measurements of electroweak
parameters such as the masses of the $W$ \cite{mw} and $Z$ \cite{mz} bosons, and of the top
quark \cite{mt}.

While the observed properties of the new resonance are consistent with those of the SM Higgs boson,
further measurements are required to determine if it has one of the key properties predicted by the
SM: couplings to fermions that are proportional to their masses.  Fortunately, a Higgs boson with a
mass of 125 GeV provides a wealth of decay modes in which to study its couplings.  In addition to
the subleading decays that dominate the sensitivity of the initial observation
($H\rightarrow \gamma\gamma$, $H\rightarrow ZZ$, and $H\rightarrow WW$), the leading decays
$H\rightarrow b\bar{b}$ and $H\rightarrow \tau\tau$ can be observed in various production processes
\cite{tthbb,jetstructure,multivariate,hbbvbf}.  As a result, a wide variety of Higgs boson cross
section measurements will provide incisive tests of specific SM couplings \cite{couplings1,couplings2}.

The prospects for Higgs boson discovery and measurement have been studied extensively \cite{djouadi};
however, low-rate processes observable with the full LHC design luminosity have not been completely
explored.  We investigate the sensitivity of 100 fb$^{-1}$ of $\sqrt{s} = 14$ TeV LHC data to the
Higgs boson production processes $WH$, $ZH$, and $t\bar{t}H$, followed by $H\rightarrow \tau\tau$
and at least one $W\rightarrow l\nu$ or $Z\rightarrow ll$ decay \cite{ichep}.  The potential measurement
sensitivity to the $WH$ process has been considered only perfunctorily \cite{snowmass}, though the CMS
experiment has recently performed the first $WH$ search in the $H\rightarrow \tau\tau$ decay channel at
the LHC \cite{newcms}.  Studies of $t\bar{t}H$ production have been performed in various top-quark and
tau-lepton decay channels \cite{ttH, ttH2}; we revisit $t\bar{t}+H(\rightarrow \tau\tau)$ production in
light of the demonstrated performance of the ATLAS and CMS experiments in separating hadronic tau decays
from the large hadronic jet background in data \cite{tauperf}.  Combining the prospects for measurements
of associated Higgs boson production in the $\tau\tau$ decay channel with those in the $b\bar{b}$ decay
channel~\cite{tthbb,jetstructure, multivariate} improves the expected LHC sensitivity to the Yukawa
coupling ratio $g_{Hbb}/g_{H\tau\tau}$.  This ratio is determined at tree level by the bottom-quark and
tau-lepton masses and is thus sensitive to differences in the source of mass for quarks and
leptons \cite{btotaucoupling}.  The ratios of associated Higgs production measurements also directly
provide the coupling ratios $g_{Htt}/g_{HWW}$ and $g_{HWW}/g_{HZZ}$.

This paper is structured as follows: Section~\ref{sec:strategy} outlines the procedures for generating,
simulating and selecting Higgs boson and background events; Section~\ref{sec:analysis} describes the
specific selection and expected signal and background yields for the $WH$, $ZH$, and $t\bar{t}H$
processes; Section~\ref{sec:results} presents the results of the fit to cross section in each channel
and the uncertainties on partial-width ratios; and Section~\ref{sec:conclusions} summarizes our
conclusions.

\section{Signal and background simulation}
\label{sec:strategy}
We simulate all signal and background processes using the {\sc sherpa} \cite{sherpa}
event generator, except for $W + 6~{\rm jets}$, which is simulated using
{\sc alpgen} \cite{alpgen} for the hard process and {\sc pythia} \cite{pythia} for
the hadronization and showering.   The $W + 6~{\rm jets}$ and $t\bar{t} + 2~{\rm jets}$
cross sections are obtained from {\sc alpgen}; all other processes are normalized to
cross sections calculated at next-to-leading order in $\alpha_s$.  Detector resolutions and 
efficiencies are modelled using the {\sc delphes} simulation \cite{delphes} with corrections 
based predominantly on ATLAS \cite{atlascsc} performance projections; similar performance is 
expected with the CMS \cite{cmstdr} detector.  Events are selected using the reconstructed 
{\sc delphes} objects.

\subsection{Event generation and cross sections}
We use CTEQ6M parton distribution functions \cite{cteq} for cross section calculations.
Samples are generated with quark and gluon jets included to leading order at the
matrix-element level, and additional jets modelled by parton showering.  Tau leptons
are decayed within {\sc sherpa}.

The cross sections and branching ratios for the Higgs boson production and decay processes are
shown in Table~\ref{tbl:s_xsec}.  We study Higgs boson masses ($m_H$) in the 115-135 GeV range
to investigate the dependence of the expected sensitivity on $m_H$.  Cross sections for $W/Z + H$
production are calculated with {\sc v2hv} \cite{v2hv} and include QCD corrections at NLO.  The
next-to-next-to-leading order (NNLO) QCD \cite{wzhqcd} and NLO electroweak \cite{wzhewk}
corrections are $\lesssim 5\%$ relative to the {\sc v2hv} calculation.  Cross sections for
$t\bar{t}H$ production include QCD corrections at NLO \cite{tthqcd}.  The uncertainties on
all signal cross sections are ${\cal{O}}$(10\%), while those on the branching ratios
determined from {\sc hdecay} \cite{hdecay} are ${\cal{O}}$(1\%).

\begin{table*}[!ht]
\begin{tabular}{ccccc}
\hline
\hline
$m_H$ (GeV) & $\sigma(pp\rightarrow WH)$ & $\sigma(pp\rightarrow ZH)$ &
$\sigma(pp\rightarrow t\bar{t}H)$ & BR($H \rightarrow \tau\tau$) \\
\hline
115 & 1.98 pb & 1.05 pb  & 0.785 pb & 0.0739 \\
120 & 1.74 pb & 0.922 pb & 0.694 pb & 0.0689 \\
125 & 1.53 pb & 0.813 pb & 0.623 pb & 0.0620 \\
130 & 1.35 pb & 0.718 pb & 0.559 pb & 0.0537 \\
135 & 1.19 pb & 0.638 pb & 0.501 pb & 0.0444 \\
\hline
\hline
\end{tabular}
\caption{Higgs boson production cross sections \cite{v2hv, tthqcd} and branching ratios \cite{hdecay}
as a function of $m_H$.}
\label{tbl:s_xsec}
\end{table*}

The dominant backgrounds to the $W/Z + H$ processes are the production of dibosons, where the
bosons decay leptonically, and $W/Z$ + hadronic jet(s), $t\bar{t}$, and $tW$, where at least one
jet is (mis)reconstructed as a lepton.  Background production cross sections are obtained from
{\sc mcfm} \cite{mcfm} and, for $t\bar{t}$, an NLO plus NLL calculation \cite{ttbar}.  For the
$W/Z + {\rm jets}$ backgrounds, we calculate cross sections requiring the boson mass to be between
20 and 200 GeV, the jets to have $p_T > 15$ GeV and $|\eta| < 3.5$, and, when there are two or more
jets, $m_{jj} > 20$ GeV.  The cross sections multiplied by SM branching ratios \cite{pdg} are shown
in Table~\ref{tbl:b_xsec}.

The $t\bar{t}H$ process, with $H\rightarrow \tau\tau$, has relatively little background.  The
irreducible background $t\bar{t}Z$ has a cross section \cite{ttznlo} that is lower than the signal
process.  The background where hadronic jets are (mis)reconstructed as leptons results predominantly
from $t\bar{t}$ production in association with 2 or 3 jets.  We estimate this background using a
leading-order cross section calculated with {\sc alpgen} \cite{alpgen}.  The calculation requires
jets with $p_T > 15$ GeV and $|\eta| < 3.5$, and $\Delta R > 0.7$ between jets.  The potential
background of $W + 6~{\rm jets}$ production is studied using an {\sc alpgen} cross section with
the above jet requirements and the $W$ boson mass between 50 and 120 GeV.  We find it to be
negligible.

\begin{table}[!ht]
\begin{tabular}{cc}
\hline
\hline
Production process & Cross section $\times$ BR \\
\hline
$W(\rightarrow l \nu)Z/\gamma^*(\rightarrow ll)$ & $52.4~{\rm pb} \times 3.27\% = 1.56~{\rm pb}$ \\
$Z/\gamma^{*}(\rightarrow l l)Z/\gamma^*(\rightarrow \tau \tau)$ & $17.7~{\rm pb} \times 0.340\% = 
60.2~{\rm fb}$ \\
\hline
$W(\rightarrow l \nu) + 2~\mathrm{jets}$ & $26772~{\rm pb} \times 32.4\% = 8674~{\rm pb}$  \\
$Z/\gamma^{*}(\rightarrow l l) + 1~\mathrm{jet}$ & $24466~{\rm pb} \times 10.1\% = 2471~{\rm pb}$ \\
$Z/\gamma^{*}(\rightarrow l l) + 2~\mathrm{jets}$ & $9018~{\rm pb} \times 10.1\% = 911~{\rm pb}$  \\
$W(\rightarrow l \nu) + 6~\mathrm{jets}$ & $23.5~{\rm pb} \times 32.4\% = 7.61~{\rm pb}$  \\
\hline
$t\bar{t}(\rightarrow l \nu l \nu b \bar{b})$ & $933~{\rm pb} \times 10.5\% = 97.9~{\rm pb}$ \\
$t\bar{t}(\rightarrow l \nu q \bar{q} b \bar{b}) + 2~\mathrm{jets}$ &
$255~{\rm pb} \times 43.8\% = 112~{\rm pb}$ \\
$t\bar{t}(\rightarrow l \nu l \nu b \bar{b}) + 2~\mathrm{jets}$ &
$255~{\rm pb} \times 10.5\% = 26.8~{\rm pb}$ \\
$tW(\rightarrow l \nu b l \nu)$ & $61.8~{\rm pb} \times 10.5\% = 6.49~{\rm pb}$ \\
$t\bar{t}(\rightarrow l \nu q \bar{q} b \bar{b}) Z/\gamma^{*}(\rightarrow ll)$ &
$973~{\rm fb} \times 4.34\% = 42.2~{\rm fb}$ \\
\hline
\hline
\end{tabular}
\caption{Background production cross sections obtained from {\sc alpgen} \cite{alpgen}
($W + 6$ jets and $t\bar{t} + 2$ jets), an NLO plus NLL calculation ($t\bar{t}$ \cite{ttbar}),
and {\sc mcfm} \cite{mcfm} (the remaining processes), multiplied by SM branching ratios \cite{pdg}.
In this table $l$ represents $e, \mu$ or $\tau$.}
   \label{tbl:b_xsec}
\end{table}

\subsection{Detector simulation}
\label{sec:detector}
We model detector acceptance and response using the {\sc delphes} simulation program
\cite{delphes}.  The detector consists of a charged particle tracker covering
$|\eta| < 2.5$ surrounded by a calorimeter with coverage to $|\eta| = 4.9$.  The calorimeter
has a granularity of $\Delta \eta \times \Delta \phi = 0.1 \times 0.1$ and is divided into
central ($|\eta| < 1.7$), forward ($1.7 < |\eta| < 3.2$), and endcap ($3.2 < |\eta| < 4.9$)
regions with separate resolutions.  Additional segmentation into electromagnetic (EM) and
hadronic (Had) calorimeters provides improved resolution for electrons and photons relative
to hadrons.

Detector resolutions are modelled by smearing the reconstructed momentum with a Gaussian
resolution.  Muon resolution is parameterized as $\sigma(p_T)/p_T = 1\%$, which is the approximate
expected resolution of muons from weak boson decays \cite{cmsmu}.  Calorimeter resolutions are
parameterized as

\begin{displaymath}
\frac{\sigma_{E}}{E} = C \oplus \frac{S}{\sqrt{E}} \oplus \frac{N}{E},
\end{displaymath}

\noindent
where $E$ is expressed in units of GeV.  In the central and forward EM calorimeters the only
non-negligible term applied is a sampling term $S$ of about $10\% \sqrt{{\rm GeV}}$.  The
resolution of the hadronic calorimeters is also dominated by the sampling term, which ranges
from about 50\% in the central region to $\approx 95\%$ in the endcap region.  The constant
terms $C$ provide small additional contributions of about 3\% and 7.5\% in the central and
endcap regions respectively.  The sampling and constant terms in the forward region are
roughly in the middle of the corresponding central and endcap terms.  The noise term ($N/E$)
is negligible for the final states we consider.

The detector acceptance for electrons, taus, and charged-particle tracks is assumed to be
$|\eta| < 2.5$.  Muon coverage is assumed to extend to $|\eta| < 2.7$.  Because of the potential
challenges in reconstructing jets in the forward region at high luminosity, we conservatively
assume a jet acceptance of $|\eta| < 3.5$.  Jets are reconstructed with the anti-$k_t$
algorithm \cite{antikt} with cone radius 0.4 and, if $|\eta| < 2.5$, are identified as
originating from either a $b$ quark or a lighter quark or gluon.  Hadronic tau decays are
identified as jets with $>90\%$ of their energy within a cone of $\Delta R < 0.15$ and only
one reconstructed track with $p_T > 2$ GeV and $\Delta R < 0.4$ from the jet axis.  Electrons
and muons are identified if no additional track with $p_T > 2$ GeV lies within a cone of
$\Delta R < 0.2$ from the $e$ or $\mu$.  Finally, the $p_T$ imbalance in the event (\met) is
derived by summing over the momentum of each calorimeter tower and muon.  Muons deposit no
energy in the calorimeter in {\sc delphes}.

Efficiencies are applied to leptons according to the expected ATLAS performance \cite{atlascsc}
or, for $\tau$ identification, the ATLAS detector performance from 2011 data \cite{tauperf}
(Table~\ref{tab:id}).  Trigger efficiencies are based on a trigger requiring a single electron
or muon with $p_T > 25$ GeV.  While actual thresholds may be higher, the presence of multiple
leptons should allow a set of triggers with a similar combined efficiency.  Rates for hadronic
jets to be misidentified as leptons are also based on expected ATLAS performance and are shown in
Table~\ref{tab:id}.  Since we always consider electrons and muons together, the averages of $e$
and $\mu$ efficiencies and misidentification rates are the relevant quantities (rather than the
individual rates).

\begin{table}[!t]
\begin{tabular}{cccc}
\hline
\hline
Object & \multicolumn{2}{c}{Efficiency (\%)} & Misidentification rate (\%)\\
       & Trigger & Identification & \\
\cline{1-4}
%$e$        & 94.3 & 64.2 & 0.0108 \\
%$\mu$      & 80.0 & 94.2 & 0.169  \\
$e/\mu$     & 87   & 79 & 0.085 \\
$\tau_{h}$ & -     & 30 & 1.0   \\
\hline
\hline
\end{tabular}
\caption{Lepton trigger and identification efficiencies, and rates for hadronic jets to be
misidentified as leptons.  Efficiencies and misidentification rates are applied to objects
at the generator level. }
\label{tab:id}
\end{table}

The \met resolution is expected to degrade from additional interactions present at the design
luminosity of $\cal{L} = ~\mathrm{10^{-34}~cm^{-2}~s^{-1}}$.  At this luminosity and 25 ns
bunch spacing, one can expect $\approx 25$ interactions per crossing.  Each interaction
will deposit $\sum E_{T} \approx 30$ GeV in the calorimeter, and the \met\, resolution is
expected to be $\approx 0.5 \sqrt{\sum E_{T}}$ \cite{atlascsc}.  To account for the
degradation in \met\, resolution from the additional interactions, we add a Gaussian
resolution with $\sigma = 15$ GeV to the projections $\displaystyle{\not}p_x$ and
$\displaystyle{\not}p_{y}$.

The performance of $\tau_h$ identification at $\sqrt{s} = 14$ TeV in the presence of 25
additional interactions is difficult to predict.  In addition to our nominal efficiency
of 30\%, we study an optimistic scenario where the efficiency is increased to 40\% for
the same misidentification rate.  The two scenarios give an indication of the effect of
the performance of tau identification on the results.

\section{Event selection}
\label{sec:analysis}
Each of the three production channels ($WH$, $ZH$ and $t\bar{t}H$) is subdivided according to the 
decay of the tau leptons originating from the Higgs boson.  The general strategy is to define a 
simple cut-based selection for each decay channel and then to perform a one-dimensional likelihood 
fit to a mass-based distribution.  The simple selection limits the number of assumptions on the 
detector performance; the key assumptions are relatively low jet-to-$\tau_h$ misreconstruction 
rates and reasonable \met resolution.  The fit reduces the effect of normalization uncertainties 
on the background.  We assume that the dominant uncertainties will result from extrapolations of 
control regions in data, and will not significantly affect the sensitivity.

\subsection{$WH$ selection}
\label{sec:WH}
Considering only the leptonic $W$-boson decays, the $WH$ final state contains one lepton,
\met from the neutrino, and two tau leptons from the Higgs boson decay.  Events where at
least two $\tau$ leptons decay hadronically are not included in this study because the
$\approx 1\%$ jet-to-$\tau_h$ misidentification rate leads to overwhelming background
from $W~\mathrm{+~jets}$ production.  Events where all tau leptons decay leptonically
are also not included because the relatively low branching ratio results in marginal
sensitivity in the corresponding final state; adding it to the final state with one
$\tau_h$ would reduce the uncertainty on the $WH$ cross section by $\approx 20$\%.  We
study the final state $l_W\tau_l\tau_h\met$, where $l_W$ is an $e$ or $\mu$ assumed to
come from a $W$-boson decay and $\tau_l$ is an $e$ or $\mu$ assumed to come from a
tau-lepton decay.  We define $l_W$ by $p_T^{l_W} > p_T^{\tau_l}$; in more than 80\%
of signal events the lepton from the $W$ boson decay has higher $p_T$ than that from
the tau lepton decay.

\begin{figure}[!ht]
\begin{center}
\includegraphics[width=0.47\textwidth]{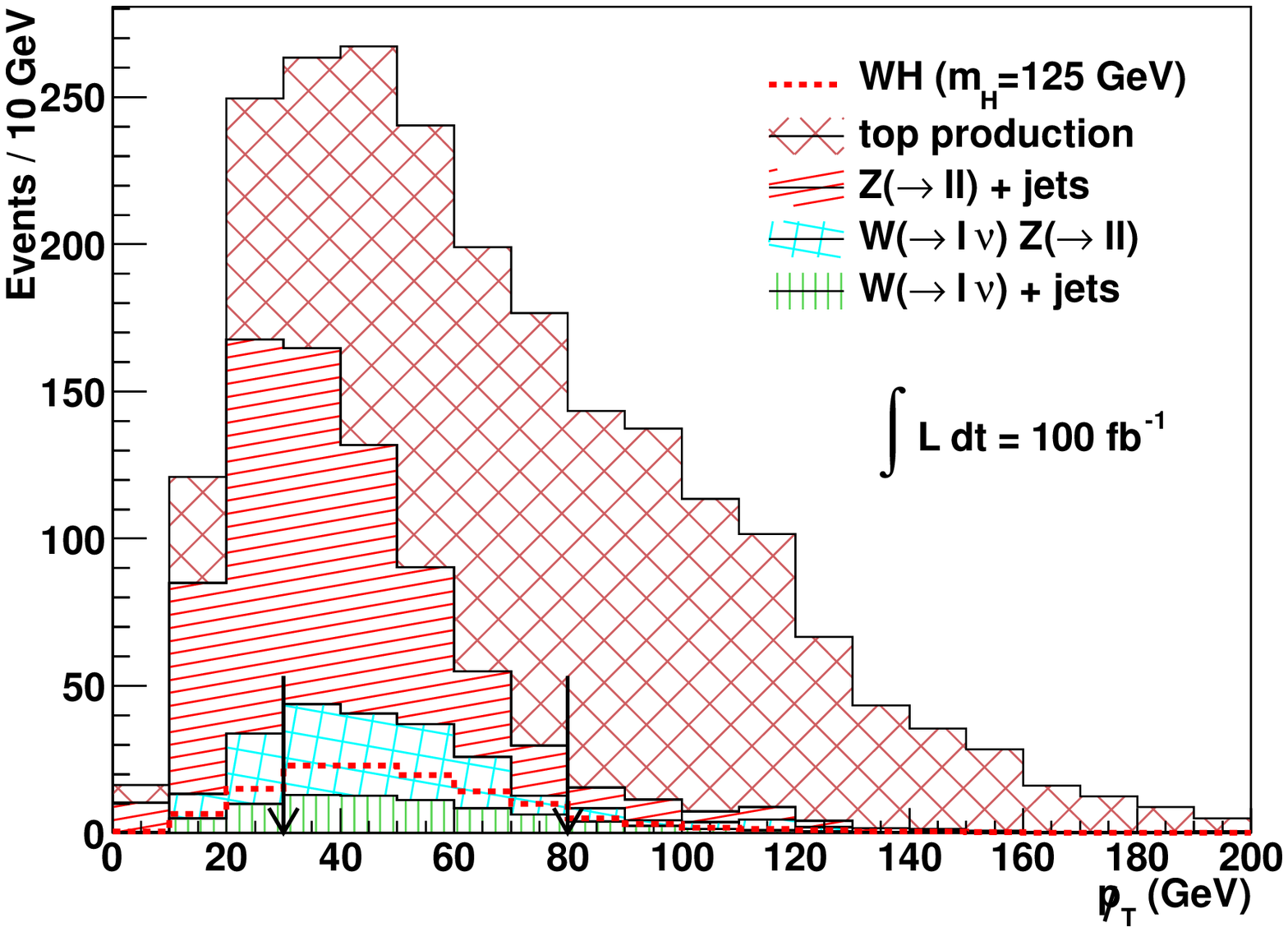}
\includegraphics[width=0.47\textwidth]{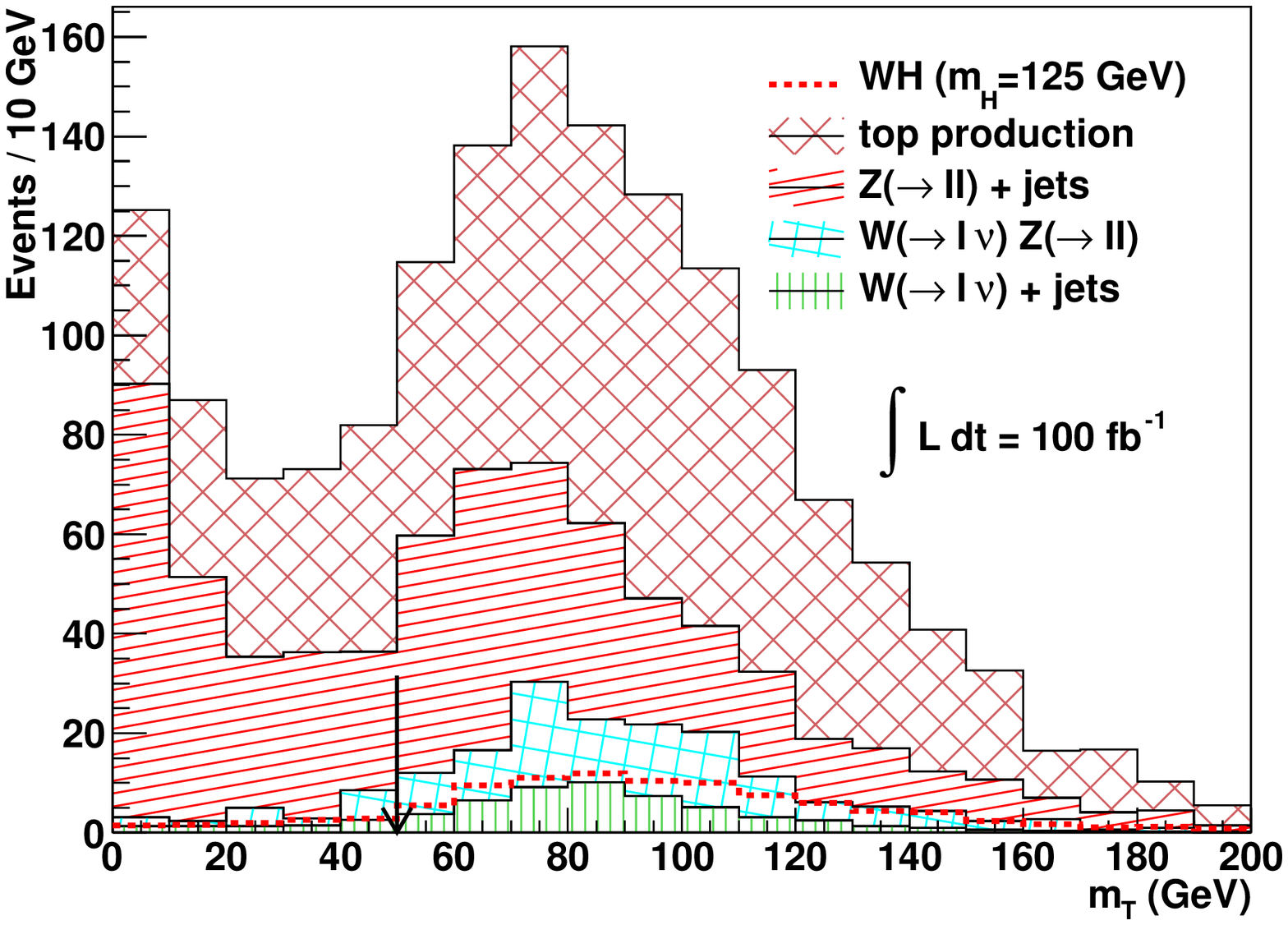}
\includegraphics[width=0.47\textwidth]{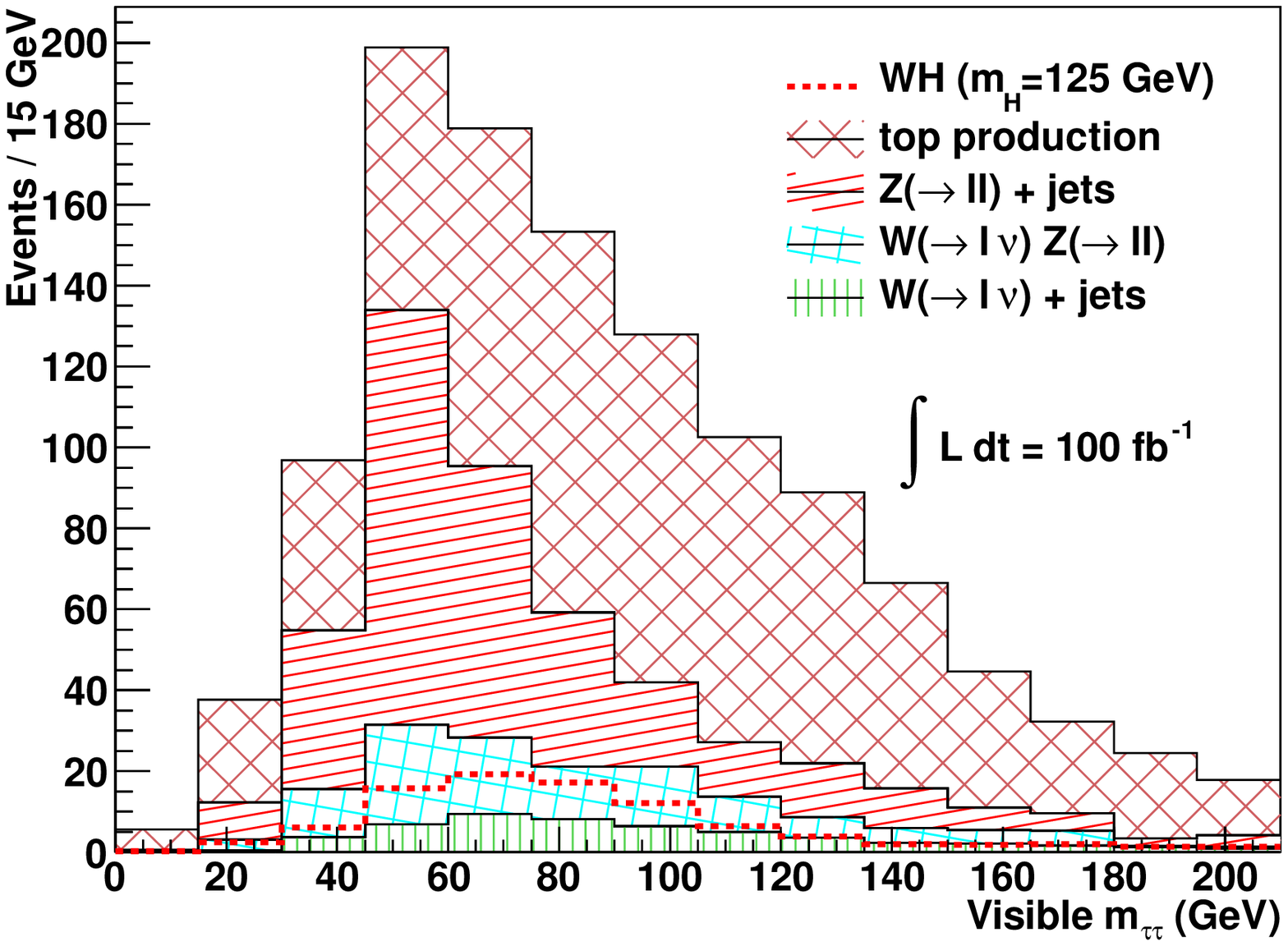}
\caption{The $\met$ (top), $m_T$ (middle) and $m(\tau_h\tau_l)$ distributions after all
$l_W\tau_l\tau_h~\met$ selection requirements, except for requirements on the plotted
distribution.  The selected regions are between the arrows in the $\met$ plot and above
the arrow in the $m_T$ plot.  Shown are the $WH$ signal (dashed line) and the following
backgrounds:  top-quark (diagonal-hatched region), $Z~+$ jet (tilted-hatched region),
$WZ$ (tilted-lined region), and $W~+$ jet (horizontal-lined region) production.}
\label{fig:WH_n_1}
\end{center}
\end{figure}

There are several background contributions to the $l_W\tau_l\tau_h$\met final state.
Production of $W$ and $Z$ bosons in association with hadronic jets, as well as $t\bar{t}$
and $tW$ decays, contribute when at least one hadronic jet is misreconstructed as a
lepton.  We model these backgrounds by applying the hadronic misidentification rates
listed in Table~\ref{tab:id} to all jets in the events.  Production of $WZ$ background and
$WH$ signal are modelled using MC acceptances, with corrections for trigger and identification
efficiencies (Table~\ref{tab:id}).

The presence of neutrinos from the $\tau$-lepton and $W$-boson decays prevents a full
reconstruction of the Higgs boson mass.  However, the ``visible mass,'' defined as the
invariant mass of the $\tau_l\tau_h$ pair, is correlated with the Higgs boson mass.
We perform a likelihood fit to the visible mass distribution to extract the signal yield.

\begin{table*}[!tbp]
\centering
\begin{tabular}{cccc}
\hline
\hline
Selection & ~$N^{WH}_s$~ & ~$N^{WH}_b$~ & $N^{WH}_s / \sqrt{N^{WH}_b}$~ \\
\hline
$p_T^{l_W} > 25$ GeV, $p_T^{\tau_l,\tau_h} > 15$ GeV, $\sum q_l = \pm 1$ and no jet &
233 &  171408 &  0.6 \\
$30 < \met < 80$ GeV  &  137 &  19124 &  1.0 \\
$m_T > 50$ GeV &  103 &  1582 &  2.6 \\
No opposite-sign same-flavour $l_W\tau_l$ &  92 &  1177 &  2.7 \\
\hline
\hline
\end{tabular}
\caption{The numbers of $WH$ signal and background events passing each set of requirements,
for an integrated luminosity of 100 fb$^{-1}$ and $m_H = 125$ GeV.  Also shown is the signal
over the square root of background, a measure of the statistical sensitivity to the signal.
Additional sensitivity is gained from a fit to the visible mass distribution. }
\label{tab:WHllh}
\end{table*}

Event selection begins with the reconstructed objects in the final state.  For the signal
process, an $e$ or $\mu$ from the $W$-boson decay typically has the highest $p_T$ of the
three charged leptons, with a $p_T$ distribution that peaks around 40 GeV.  We therefore
require $p_T^{l_W} > 25$ GeV.  The unobserved neutrinos in tau-lepton decays reduce the $p_T$
of the reconstructed objects, so a $p_T$ threshold of 15 GeV is applied to $\tau_l$ and
$\tau_h$.  Background from $W +$ jets is suppressed by requiring the charges of the leptons
($q_l$) to sum to $\pm 1$.  Events are required to have no jet with $p_T > 25$ GeV and
$|\eta| < 3.5$, reducing both top-quark and $W/Z$ + jet(s) backgrounds.  A requirement of
$\met > 30$ GeV reduces background from $Z +$ jet production, and an upper bound of
$\met < 80$ GeV reduces top-quark background.

\begin{table}[!tbp]
\centering
\begin{tabular}{cc}
\hline
\hline
Process & Number of events \\
\hline
$t\bar{t}(\rightarrow l \nu l \nu b \bar{b})$ & 573 \\
$Z/\gamma^{*}(\rightarrow l l) + 1~\mathrm{jet}$ & 330 \\
$tW(\rightarrow l \nu b l \nu)$ & 112 \\
$W(\rightarrow l \nu) Z/\gamma^{*}(\rightarrow \tau\tau)$  & 81 \\
$W(\rightarrow l \nu) + 2~\mathrm{jets}$  & 52 \\
$W(\rightarrow l \nu) Z/\gamma^{*}(\rightarrow ee/\mu\mu)$  & 30 \\
\hline
Total & 1177 \\
\hline
\hline
\end{tabular}
\caption{The contribution of each background to the $l_W\tau_l\tau_h\met$ final state for
an integrated luminosity of 100~fb$^{-1}$. }
\label{tab:WHbd}
\end{table}

\begin{table}[!tbp]
\centering
\begin{tabular}{ccc}
\hline
\hline
$m_H~\mathrm{(GeV)}$  & ~$N^{WH}_{s}$~ & ~$N^{WH}_{s} / \sqrt{N^{WH}_{b}}$~ \\
\hline
115 & 122 & 3.6 \\
120 & 109 & 3.2 \\
125 & 92  & 2.7 \\
130 & 70  & 2.0 \\
135 & 52  & 1.5 \\
\hline
\hline
\end{tabular}
\caption{The number of $WH$ signal events for $m_H$ in the range 115-135 GeV, and the
statistical significance of the excess of signal events over background in 100 fb$^{-1}$
of integrated luminosity. }
\label{tab:WHsig}
\end{table}

The significant background from $Z(\rightarrow \tau\tau)~+$ jet production contributes primarily
when the tau leptons decay leptonically and the jet is misreconstructed as a $\tau_h$.  The
tau lepton from the $Z$ boson decay is highly boosted and its decay products are nearly
collinear.  In a class of $Z~+$ jet events, the reconstructed \met is aligned with $l_W$,
while in signal events the $\met$ is rarely aligned with $l_W$.  Defining the transverse
mass as $m_T = \sqrt{2(p_T^{l_W}\met - p_x^{l_W} \metx - p_y^{l_W} \mety})$, we suppress
$Z~+$ jet events with the requirement $m_T > 50$ GeV.  Additional background rejection
could be achieved with a similar transverse mass requirement on $\tau_l$ and \met; however,
there would be larger reduction in signal since there are neutrinos collinear with $\tau_l$ in
signal events.

A final selection requirement of no opposite-charge, same-flavor $l_W\tau_l$ further reduces
background from $Z~+$ jet production, removing most events with $Z$ bosons decaying to $e$
or $\mu$ pairs.  Decays of $Z$ bosons to tau-lepton pairs are also reduced with this requirement,
and could be further reduced by removing events with an oppositely charged electron and muon.
However, the loss of signal from such a requirement would be relatively large, and the
statistical sensitivity would degrade.

Figure~\ref{fig:WH_n_1} shows the $\met$, $m_T$ and $m(\tau_h\tau_l)$ distributions with all
selection requirements applied, except those on the plotted quantity.  The numbers of signal
($N_s^{WH}$) and background ($N^{WH}_b$) events, as well as $N_s^{WH}/\sqrt{N^{WH}_b}$, are given
in Table~\ref{tab:WHllh} after each selection requirement.  The detailed contribution of each
background and the dependence of the signal yield on $m_H$ are shown after all selection in
Tables~\ref{tab:WHbd} and~\ref{tab:WHsig}, respectively.

The selection gives modest statistical sensitivity to $WH$ production, but the sensitivity is
improved with a fit to the visible mass distribution.  Normalization uncertaintes will be
mitigated by this fit, though uncertainties on the shape of the visible mass distribution are
also relevant; we assume the systematic uncertainties can be sufficiently constrained by studying
independent kinematic regions (for example, the high-$\met$ region for top production, and the
low-$m_T$ region for $Z~+$ jet production).

\subsection{$ZH$ selection}
\label{sec:ZH}
In contrast to $WH$ production, $ZH\rightarrow ll\tau\tau$ production is
dominated by an irreducible background ($ZZ$), with relatively low signal
statistics in 100 fb$^{-1}$ of integrated luminosity.  Thus, the selection
strategy is to apply few requirements and to combine the $l_Z l_Z\tau_h\tau_h$
and $l_Z l_Z\tau_h\tau_l$ decay channels, where $l_Z$ is an $e$ or $\mu$.
The $l_Z l_Z \tau_l\tau_l$ channel adds only marginal sensitivity because
of the small branching ratio and the increased $ZZ$ background.

\begin{table*}[!tp]
\centering
\begin{tabular}{cccc}
\hline
\hline
Selection & $N^{ZH}_s$ & $N^{ZH}_b$ & $N^{ZH}_s / \sqrt{N^{ZH}_b}$ \\
\hline
Opposite-charge $\tau_h\tau_h$ and $l_Z l_Z$; \\
highest (lowest) $p_T^{l_Z} > 25~(15)$ GeV;
$p_T^{\tau_h} > 25$ GeV & 32 & 193 & 2.3 \\
Collinear mass solution & 26 & 144 & 2.1 \\
\hline
Opposite-charge $\tau_h\tau_l$ and $l_Z l_Z$; \\
highest (lowest) $p_T^{l_Z} > 25~(15)$ GeV;
$p_T^{\tau_h(\tau_l)} > 25~(15)$ GeV  & 36  & 266  & 2.2  \\
Collinear mass solution               & 30  & 188  & 2.2  \\
\hline
\hline
\end{tabular}
\caption{The numbers of $ZH$ signal and background events passing each set of requirements,
for an integrated luminosity of 100 fb$^{-1}$ and Higgs boson mass of 125 GeV.  Also shown
is the signal over the square root of background, a measure of the statistical sensitivity
to the signal.  Additional sensitivity is gained from a fit to the collinear mass
distribution. }
\label{tab:ZH}
\end{table*}

In addition to the irreducible $ZZ\rightarrow ll\tau\tau$ background, reducible
backgrounds from $Z~+$ jets and $t\bar{t} \rightarrow l\nu l\nu b\bar{b}$
contribute when two jets are misreconstructed as hadronic tau(s) and/or light-flavor
lepton(s).  These backgrounds are modelled by applying the hadronic misidentification
rates in Table~\ref{tab:id} to MC-generated events.  Production of $ZZ$ background
and $ZH$ signal are modelled using trigger- and identification-corrected MC
acceptances (Table~\ref{tab:id}).

\begin{table}[!htbp]
\centering
\begin{tabular}{cc}
\hline
\hline
Process & Number of events \\
\hline
$Z/\gamma^*(\rightarrow ll) Z/\gamma^*(\rightarrow \tau\tau)$ & 305 \\
$Z/\gamma^{*}(\rightarrow l l) + 2~\mathrm{jets}$ & 25 \\
$t\bar{t}(\rightarrow l \nu l \nu b \bar{b})$ & 2 \\
\hline
Total & 332 \\
\hline
\hline
\end{tabular}
\caption{The contribution of each background to the $ZH$ final state for an integrated
luminosity of 100 fb$^{-1}$. }
\label{tab:ZHbd}
\end{table}

The irreducible $ZZ$ background can be separated using the invariant mass of the
tau-lepton pair.  Since the tau leptons from the Higgs boson decay are highly boosted,
their decay products are nearly collinear.  Assuming collinear tau-lepton decays, the
net neutrino momentum from each decay can be resolved.  The resulting invariant mass
of the tau-lepton pair, or ``collinear mass'', can be expressed in the $ll\tau_h\tau_l$
decay channel as $m(\tau_h,\tau_l)/\sqrt{\chi_{h}\chi_{l}}$, where $\chi_{h (l)}$ is the
fraction of tau-lepton energy taken by $\tau_h~(\tau_l)$.  The fractions $\chi_h$ and
$\chi_l$ can be solved in terms of measured quantities,
\begin{eqnarray}
\chi_{h} & = & \frac{p_x^{\tau_h}p_y^{\tau_l} - p_y^{\tau_h}p_x^{\tau_l}}{p_x^{\tau_h}p_y^{\tau_l}
+ \metx p_y^{\tau_l} - p_y^{\tau_h}p_x^{\tau_l} - \mety p_x^{\tau_l} }, \nonumber \\
\chi_{l} & = & \frac{p_x^{\tau_h}p_y^{\tau_l} - p_y^{\tau_h}p_x^{\tau_l}}{p_x^{\tau_h}p_y^{\tau_l}
+ \metx p_y^{\tau_h} - p_y^{\tau_h}p_x^{\tau_l} - \mety p_x^{\tau_h} }.
\end{eqnarray}

\noindent
For the $ll\tau_h\tau_h$ channel, $\tau_l$ is replaced by the other $\tau_h$.  We fit
the collinear mass distribution to extract the $ZH$ signal yield after initial selection
requirements.

\begin{table}[!tbp]
\centering
\begin{tabular}{ccc}
\hline
\hline
$m_H~\mathrm{(GeV)}$  & ~$N^{ZH}_{s}$~ & ~$N^{ZH}_{s} / \sqrt{N^{ZH}_{b}}$~ \\
\hline
115 & 77 & 4.2 \\
120 & 71 & 3.9 \\
125 & 56 & 3.1 \\
130 & 45 & 2.4 \\
135 & 33 & 1.8 \\
\hline
\hline
\end{tabular}
\caption{The number of $ZH$ signal events for $m_H$ in the range 115-135~GeV, and the
statistical significance of the excess of signal events over background in 100~fb$^{-1}$
of integrated luminosity. }
\label{tab:ZHsig}
\end{table}

The selection requires two opposite-charge same-flavor leptons from the $Z$ boson decay.
If an event has multiple candidate pairs, we define the pair with invariant mass closest
to $m_Z$ as the $Z$ boson candidate decay.  The highest (lowest) $p_T$ lepton from the
decay is required to have $p_T > 25~(15)$ GeV.  We then require two opposite-charge
tau-lepton decay candidates with $p_T > 25$ GeV (or $p_T > 15$ GeV for $\tau_l$).
Table~\ref{tab:ZH} shows the numbers of signal ($N_s^{ZH}$) and background ($N^{ZH}_b$)
events, as well as $N_s^{ZH}/\sqrt{N^{ZH}_b}$, in each channel after this initial selection.
The collinear mass requirement reduces the signal yield by nearly 30\%; recovering these
events with an alternative mass variable would improve the measurement.

\begin{figure}[!tpb]
\begin{center}
\includegraphics[width=0.495\textwidth]{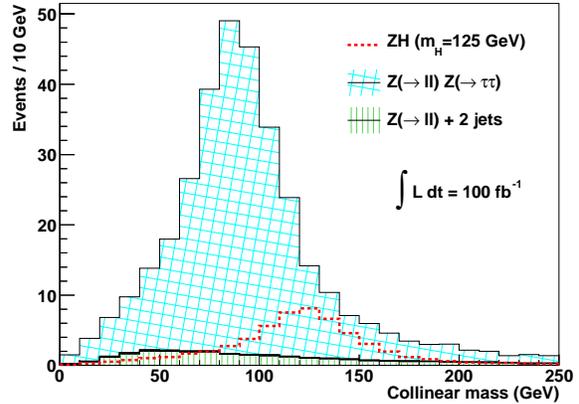}
\caption{The collinear mass distribution for the $ZH$ signal (dashed line), $ZZ$ background
(tilted-hatched region), and $Z~+$~2~jets background (vertical-lined region).  Not shown is
the negligible $t\bar{t}$ background. }
\label{fig:ZH}
\end{center}
\end{figure}

Figure~\ref{fig:ZH} shows the collinear mass distribution with all selection requirements.
The detailed contribution of each background and the dependence of the signal yield on
$m_H$ are shown after all selection requirements in Tables~\ref{tab:ZHbd} and~\ref{tab:ZHsig},
respectively.  The relatively small background and the discrimination given by the collinear
mass make the $ZH$ channel particularly promising for measuring Higgs boson decays to tau leptons.

\subsection{$t\bar{t}H$ selection}
\label{sec:ttH}
The cross section for $t\bar{t}H$ production, with the Higgs boson decaying to tau leptons,
is relatively low.  We focus on the decays with the highest branching ratios, excluding fully
hadronic $t\bar{t}$ decays because of the potentially large multijet background.  Thus we
consider $t\bar{t}\rightarrow l_W\nu q\bar{q}b\bar{b}$ and either $H\rightarrow \tau_h \tau_h$
or $H\rightarrow \tau_l \tau_h$, with $l_W$ defined by $p_T^{l_W} > p_T^{\tau_l}$.  These
final states are the same as in $WH$ production but with the addition of four jets.

For the detector performance assumed in Sec.~\ref{sec:detector}, the background is a roughly
equal mix of irreducible $t\bar{t}Z$ production and reducible $t\bar{t}~+$ jets production.
The dominant reducible background is $t\bar{t}(\rightarrow l_W \nu l\nu b\bar{b}) + 3$ jets, where
one jet is misreconstructed as a $\tau_h$, and $l$ is identified as either $\tau_h$ or $\tau_l$.
The generation of $t\bar{t} + 3$ jets at tree-level is computationally intensive; we therefore
model this background using the {\sc sherpa} $t\bar{t} + 2$ jets process, with additional jets
modelled by the {\sc sherpa} parton-showering algorithm.

\begin{figure}[!htp]
\begin{center}
\includegraphics[width=0.465\textwidth]{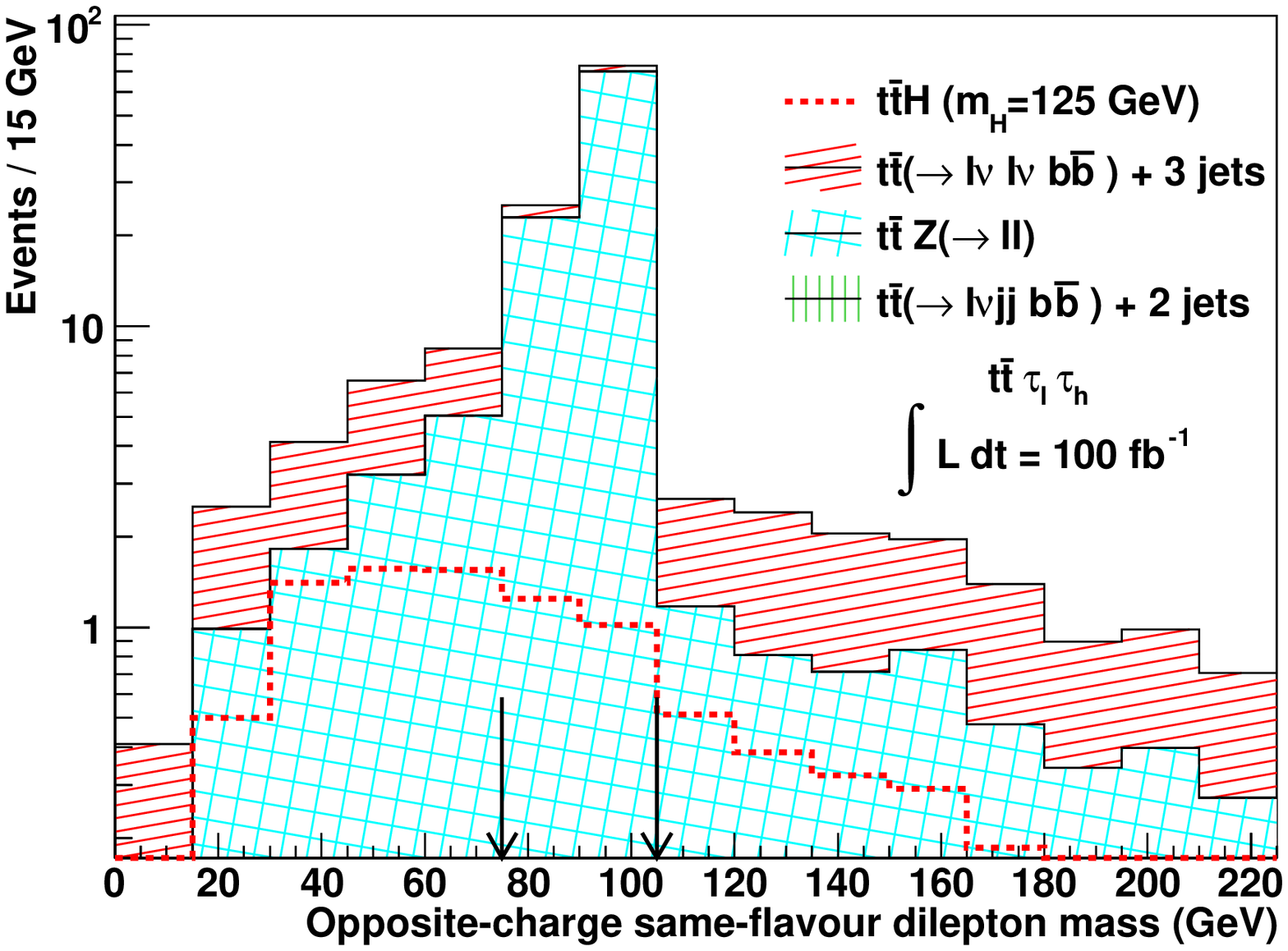}
\includegraphics[width=0.465\textwidth]{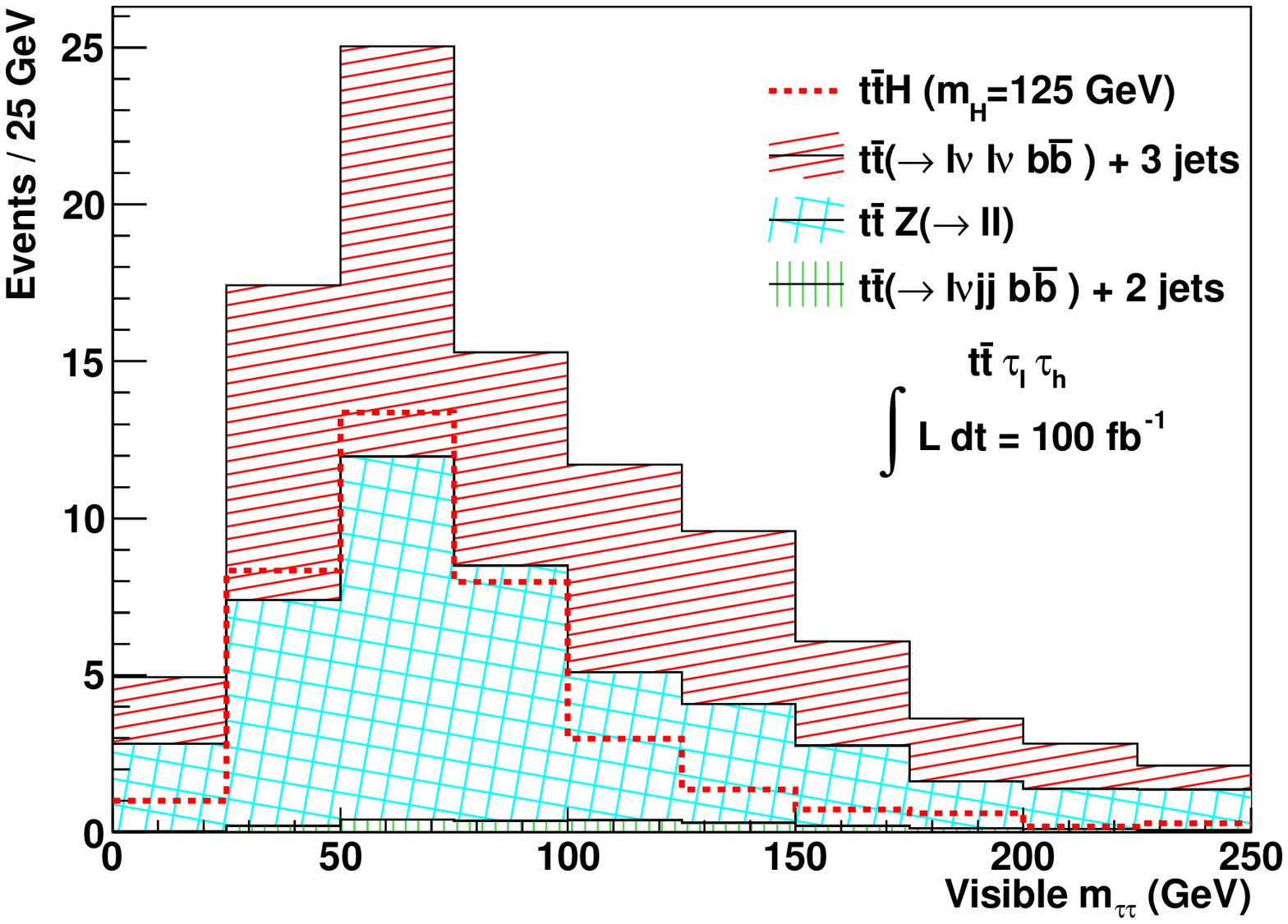}
\includegraphics[width=0.465\textwidth]{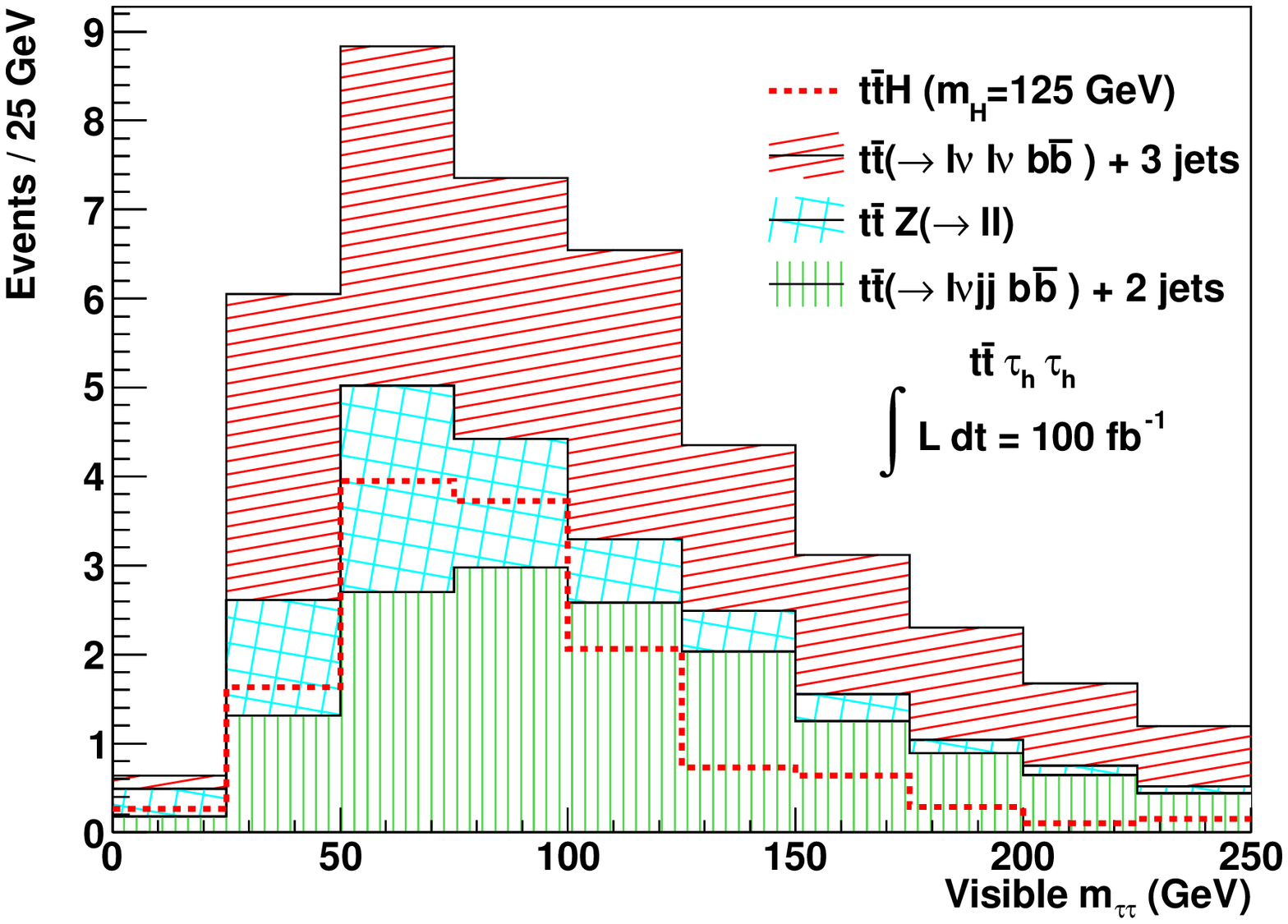}
\caption{The $m(l_W\tau_l)$ (top), $m(\tau_h\tau_l)$ (middle), and $m(\tau_h\tau_h)$
distributions after all selection requirements (except for the requirement on $m(l_W\tau_l)$
for the $m(l_W\tau_l)$ distribution).  The selected $m(l_W\tau_l)$ regions are below and
above the arrows in the $m(l_W\tau_l)$ plot.  Shown are the $ttH$ signal (dashed line) and
the following backgrounds: $t\bar{t} + 3$ jets (tilted-lined region), $tt + Z$ (tilted-hatched
region), and $t\bar{t}~+ 2$ jets (vertical-lined region) production.}
\label{fig:ttH}
\end{center}
\end{figure}

Since the reducible background consists of $t\bar{t} +$ jets, the sensitivity depends predominantly
on tau identification and the broadly peaking visible mass distribution of the tau-lepton pair.  The
irreducible $t\bar{t}Z$ background is suppressed by requiring opposite-sign same-flavor $l_W \tau_l$
pairs to have an invariant mass outside the $75-105$ GeV peak of resonant $Z$-boson production.   Other
selection requirements are $p_T^{l_W} > 25$ GeV, $p_T^{\tau_h,\tau_l} > 15$ GeV, $\sum q_\ell = \pm 1$,
and at least 4 jets.

\begin{table}[!tp]
\centering
\begin{tabular}{ccc}
\hline
\hline
Process & $t\bar{t} + \tau_h\tau_l$ & $t\bar{t} + \tau_h\tau_h$ \\
& channel & channel \\
\hline
$t\bar{t}(\rightarrow l \nu l \nu b \bar{b}) + 3$ jets                               & 52 & 20 \\
$t\bar{t}(\rightarrow l \nu q\bar{q} b \bar{b}) + Z/\gamma^*(\rightarrow ee/\mu\mu)$ & 32 & 2 \\
$t\bar{t}(\rightarrow l \nu q\bar{q} b \bar{b}) + Z/\gamma^*(\rightarrow \tau\tau)$  & 13 & 5 \\
$t\bar{t}(\rightarrow l \nu q\bar{q} b \bar{b}) + 2~\mathrm{jets}$                   & 2  & 15 \\
\hline
Total & 99 & 42 \\
\hline
\hline
\end{tabular}
\caption{The contribution of each background to the $t\bar{t}H$ final states for an integrated
luminosity of 100 fb$^{-1}$. }
\label{tab:ttHbd}
\end{table}

Figure~\ref{fig:ttH} shows the mass distribution of opposite-sign same-flavor $l_W\tau_l$ pairs and
the visible mass distributions of the tau-lepton pairs in the two decay channels 
$t\bar{t} + \tau_h \tau_l$\and $t\bar{t} + \tau_h \tau_h$.  Tables~\ref{tab:ttHbd} and~\ref{tab:ttHsig} 
respectively show the contribution of each background and the dependence of the signal yield on $m_H$ 
after all selection for both channels.  With basic object selection, reasonable sensitivity to 
$t\bar{t}H$ production can be obtained if tau leptons are identified with a similar efficiency and jet 
rejection rate to that achieved by ATLAS and CMS with $\sqrt{s} = 7$ TeV LHC data.

\begin{table}[!tbp]
\centering
\begin{tabular}{cccc}
\hline
\hline
$m_H~\mathrm{(GeV)}$  & Channel & ~$N^{ttH}_{s}$~ & ~$N^{ttH}_{s} / \sqrt{N^{ttH}_{b}}$~ \\
\hline
115 & $t\bar{t} + \tau_h\tau_l$ & 47 & 4.8 \\
    & $t\bar{t} + \tau_h\tau_h$ & 17 & 2.7 \\
\hline
120 & $t\bar{t} + \tau_h\tau_l$ & 47 & 4.8 \\
    & $t\bar{t} + \tau_h\tau_h$ & 16 & 2.5 \\
\hline
125 & $t\bar{t} + \tau_h\tau_l$ & 37 & 3.7 \\
    & $t\bar{t} + \tau_h\tau_h$ & 14 & 2.1 \\
\hline
130 & $t\bar{t} + \tau_h\tau_l$ & 30 & 3.0 \\
    & $t\bar{t} + \tau_h\tau_h$ & 11 & 1.7 \\
\hline
135 & $t\bar{t} + \tau_h\tau_l$ & 22 & 2.2 \\
    & $t\bar{t} + \tau_h\tau_h$ &  7 & 1.1 \\
\hline
\hline
\end{tabular}
\caption{The number of $ttH$ signal events in each channel for $m_H$ in the range 115-135 GeV,
and the statistical significance of the excess of signal events over background in 100 fb$^{-1}$
of integrated luminosity.  The $t\bar{t}$ pair is selected in the $l_W\nu q\bar{q}b\bar{b}$ final
state. }
\label{tab:ttHsig}
\end{table}

\section{Results}
\label{sec:results}
\begin{figure*}[!tbp]
\centering
\includegraphics[width=0.465\textwidth]{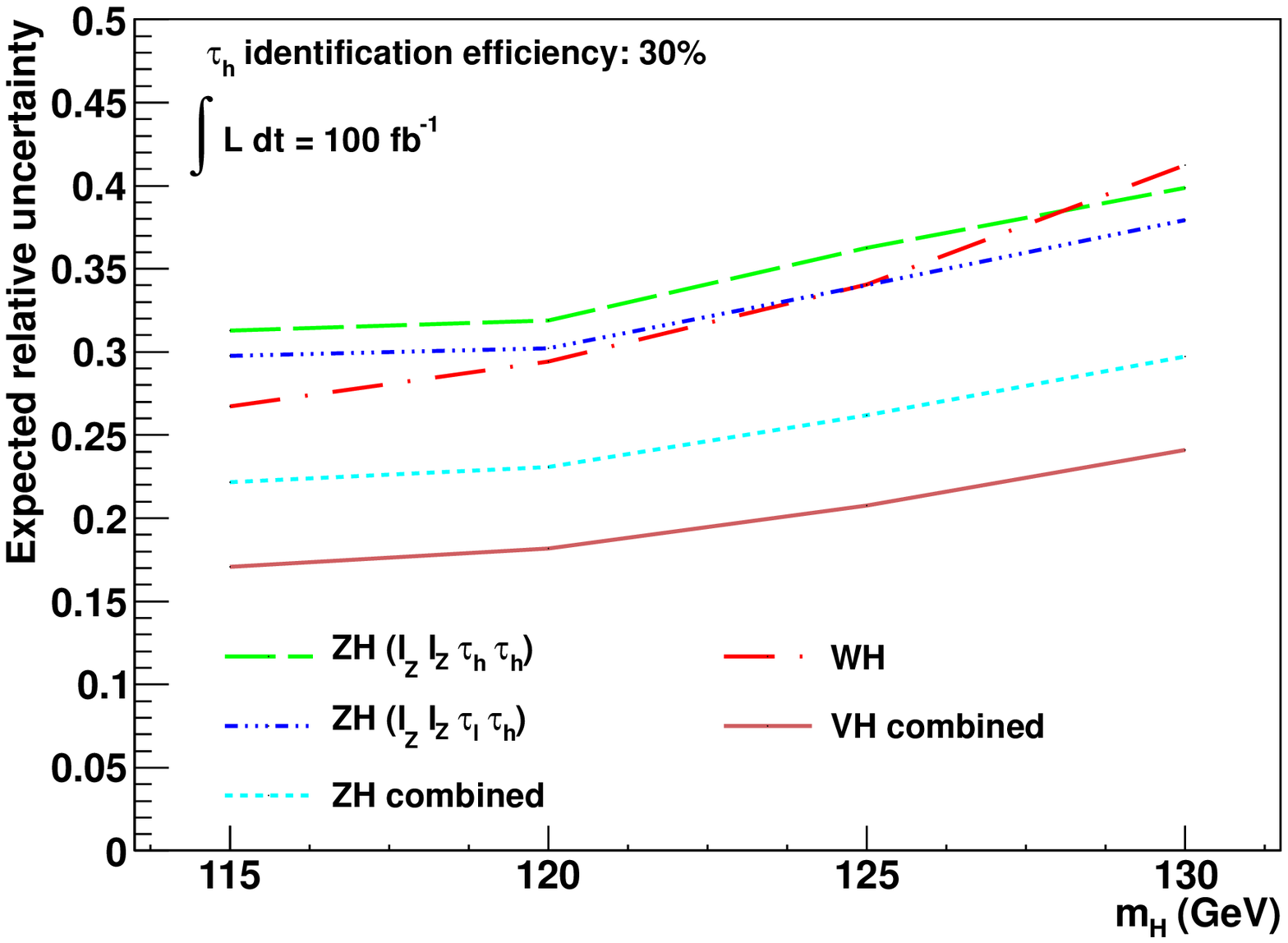}
\includegraphics[width=0.465\textwidth]{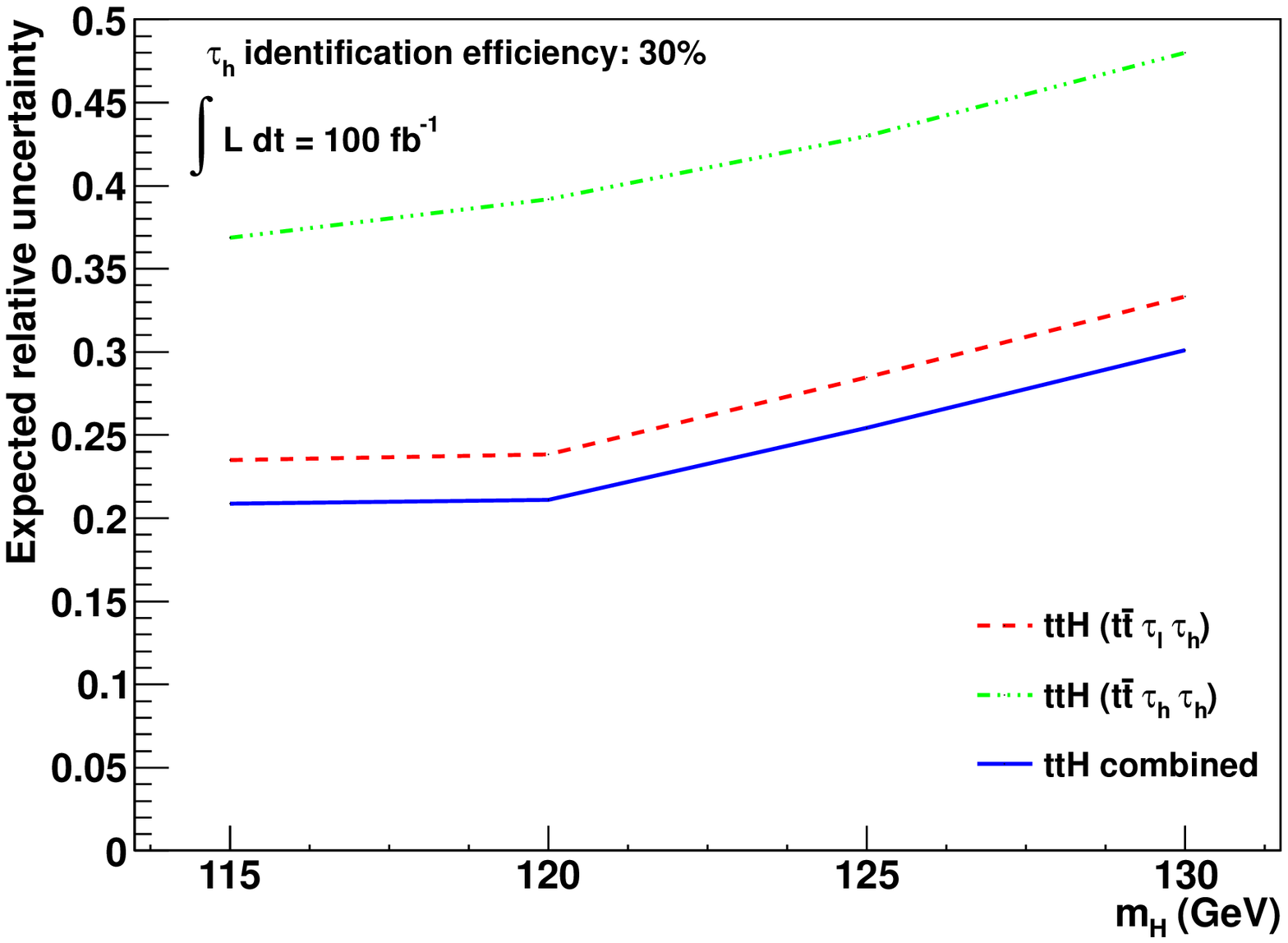}
\includegraphics[width=0.465\textwidth]{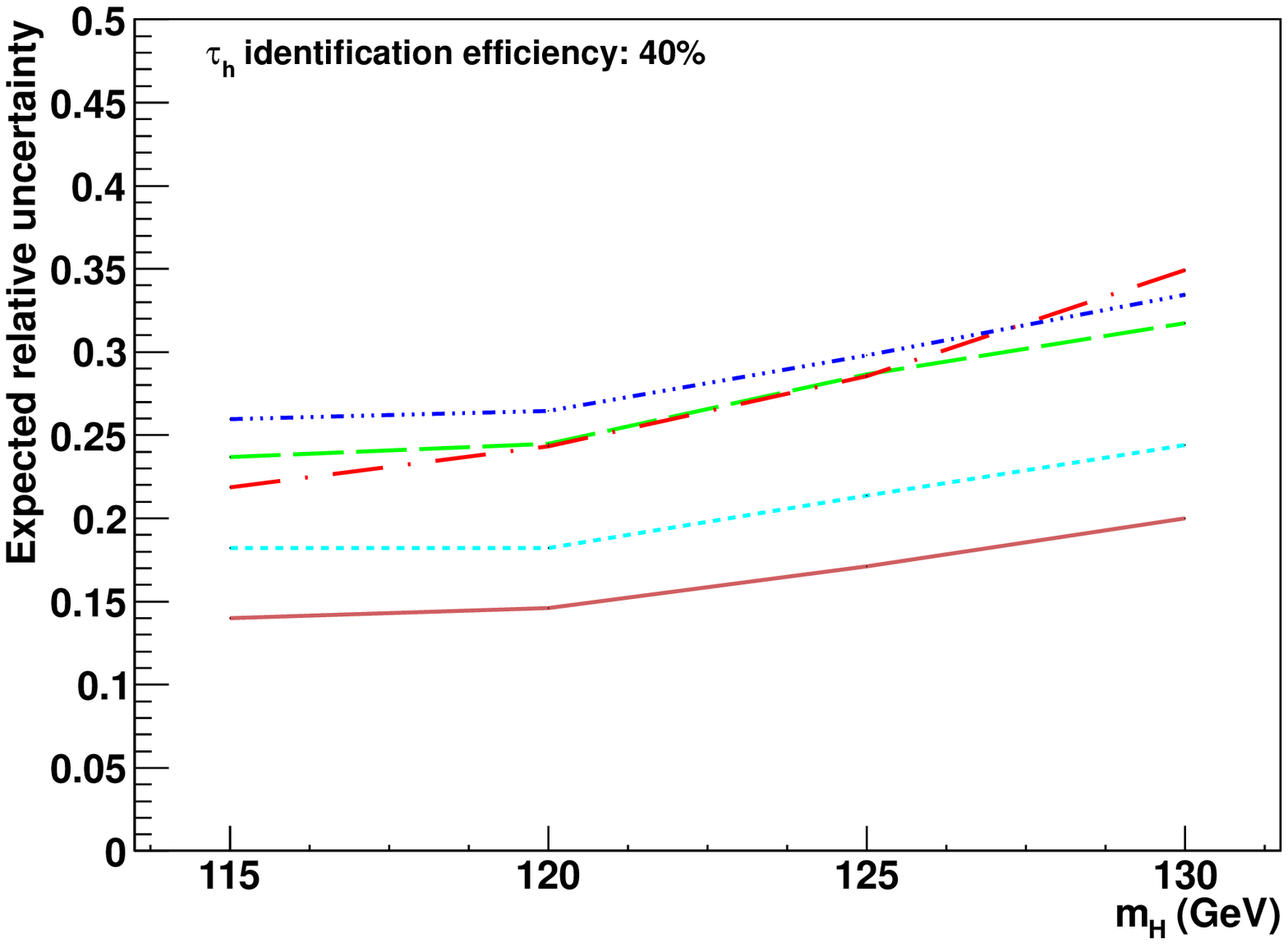}
\includegraphics[width=0.465\textwidth]{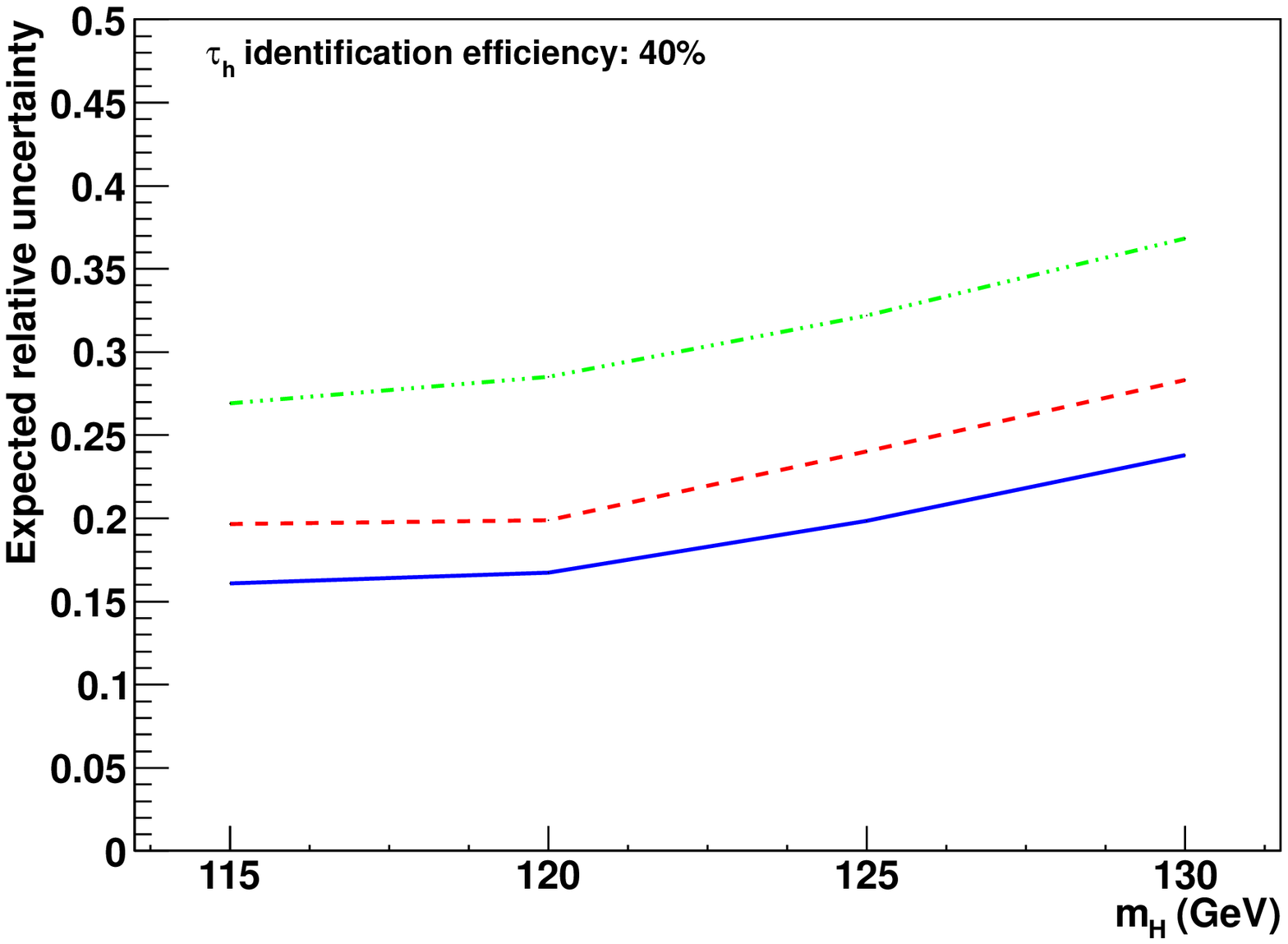}
\caption{The expected relative statistical uncertainties on $\sigma \times {\mathrm{BR}}$ of $VH$
(left, $V=W,Z$) and $t\bar{t}H$ (right) production for the nominal (top) and optimistic (bottom)
tau identification performance scenarios. }
\label{fig:dsigmabr}
\end{figure*}

e determine the expected sensitivity to the cross section of a given process using pseudoexperiments
\cite{ichep}.  In each pseudoexperiment, data are produced according to a Poisson distribution in each
bin of the relevant mass-based fit distribution, where the mean of the Poisson is equal to the combined
signal and background in that bin.  The number of signal events is determined by minimizing the negative
log likelihood of the fit distribution.  This procedure is performed for $10^4$ pseudoexperiments for
each process, and the uncertainty is taken to be the root-mean square of the resulting signal-yield
distribution.  The relative statistical uncertainties on $\sigma \times {\mathrm{BR}}$ of the signal
processes are shown in Fig.~\ref{fig:dsigmabr}.

\begin{figure}[!tbp]
\centering
\includegraphics[width=0.465\textwidth]{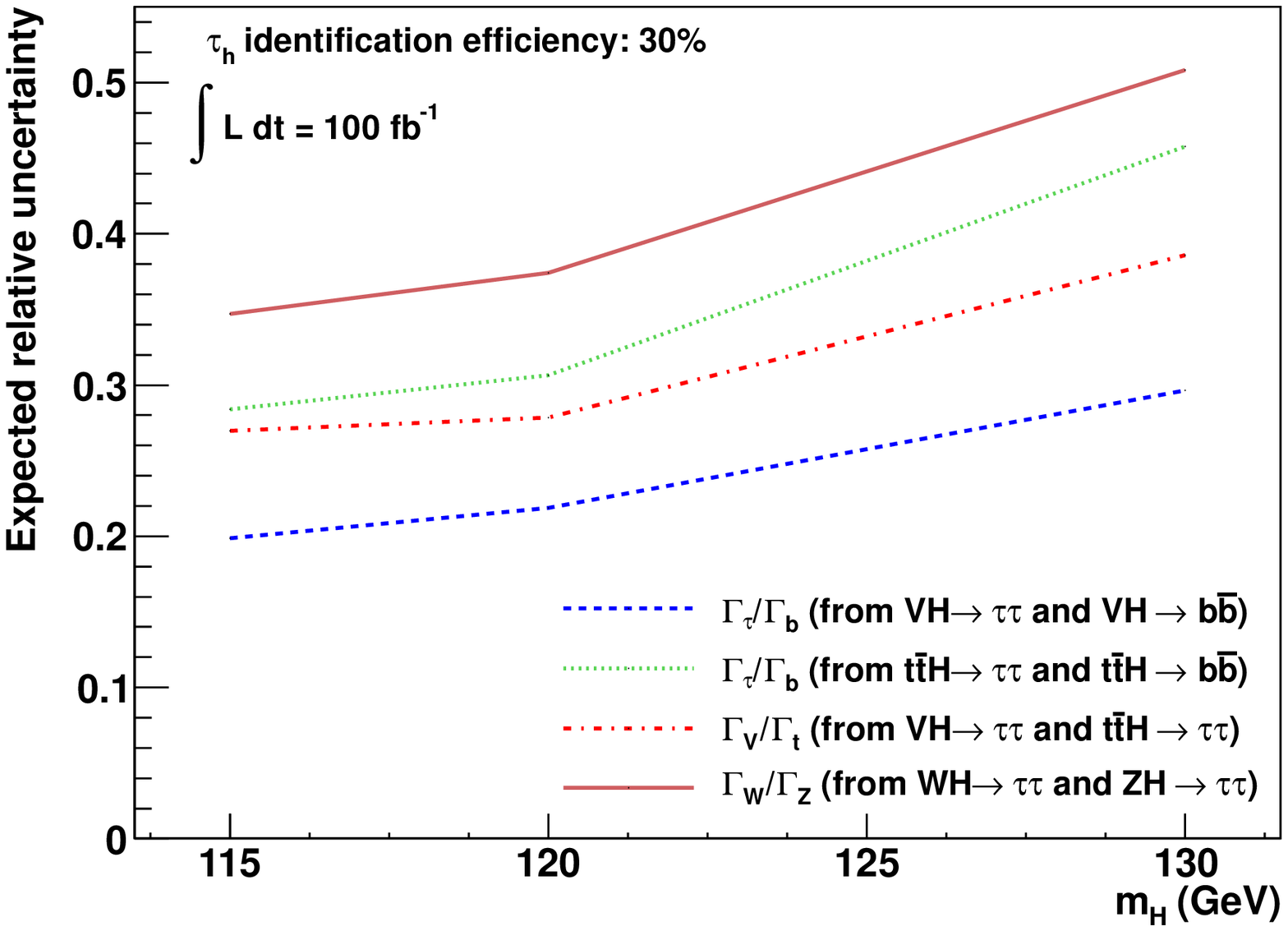}
\includegraphics[width=0.465\textwidth]{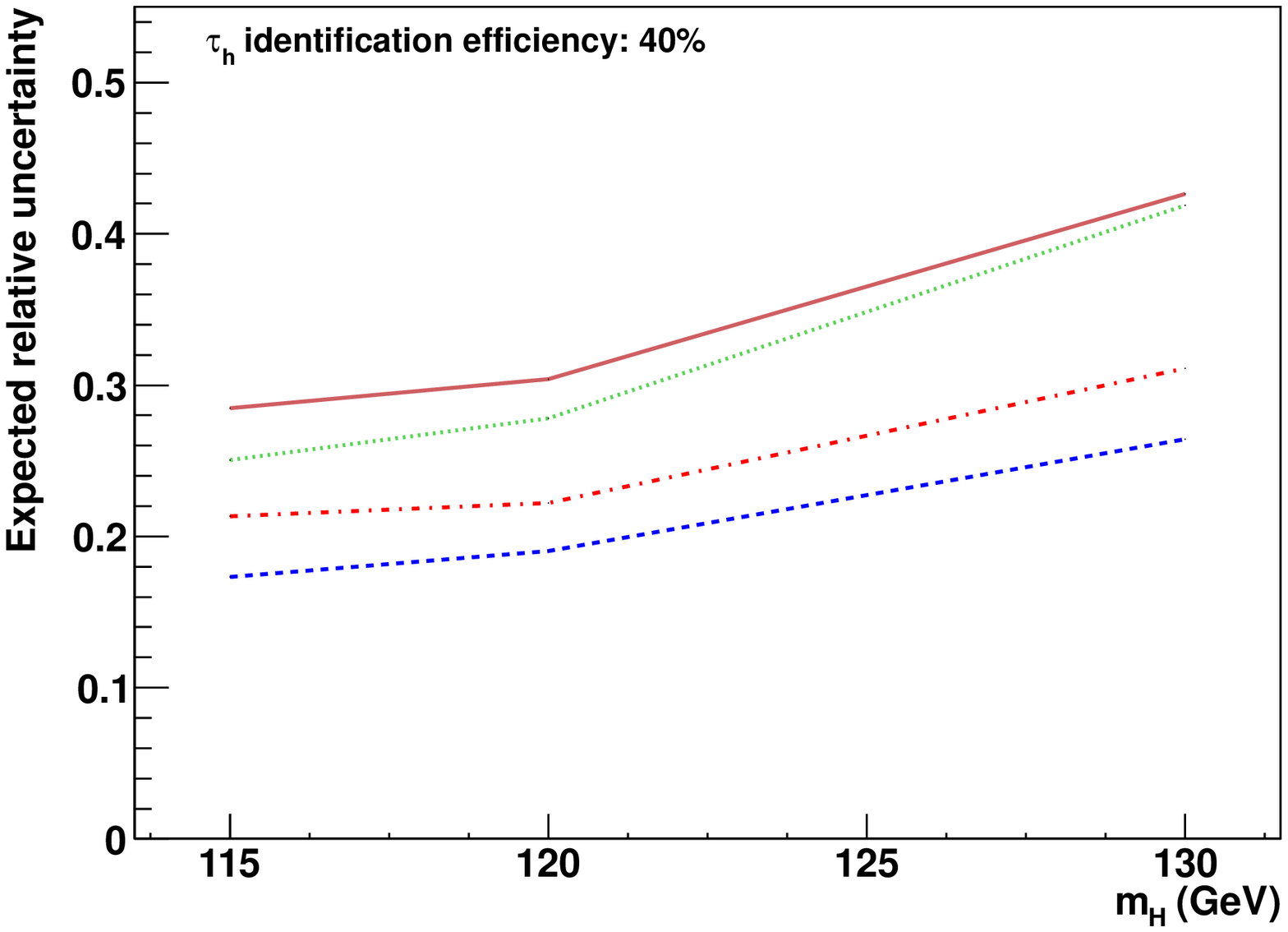}
\caption{The expected relative statistical uncertainties on ratios of partial widths using measurements
of associated Higgs boson production and decays to tau-lepton or bottom-quark pairs.  Each partial width
$\Gamma_i$ corresponds to the trilinear interaction of a Higgs boson to another particle $i$.  Shown are
the nominal (top) and optimistic (bottom) detector performance scenarios. }
\label{fig:dwidthratio}
\end{figure}

The cross section of a given signal process includes the product of partial widths for the production and
decay vertices of the Higgs boson.  Individual partial widths can be determined by taking cross-section 
ratios, providing direct access to the individual couplings of the Higgs boson to SM particles.  We 
expect this procedure to provide the additional benefit of cancelling many experimental uncertainties.  
From the ratios of cross section measurements studied in this paper, and from the expected uncertainties 
on the measurements of associated Higgs production in its decays to bottom quarks (Table~\ref{tab:hbb}), 
we obtain the expected sensitivity to partial width ratios shown in Fig.~\ref{fig:dwidthratio}.

\begin{table}[!tbp]
\centering
\begin{tabular}{ccc}
\hline
\hline
$m_H~\mathrm{(GeV)}$  & ~$VH$~ & ~$t\bar{t}H$~ \\
\hline
115 & 10\% & 19\% \\
120 & 12\% & 22\% \\
130 & 17\% & 34\% \\
\hline
\hline
\end{tabular}
\caption{The assumed relative uncertainties on $V(H\rightarrow b\bar{b})$ and
$t\bar{t}(H\rightarrow b\bar{b})$~\cite{jetstructure} cross section measurements in data
corresponding to 100 fb$^{-1}$ of integrated luminosity. }
\label{tab:hbb}
\end{table}

\section{Conclusions}
\label{sec:conclusions}
With the recent discovery of a resonance with cross sections consistent with that of the SM Higgs 
boson, tests of the specific SM predictions of the Higgs boson couplings are a high priority.  
A Higgs boson with a mass of 125 GeV can be measured in a wealth of production and decay channels.  
We have performed a detailed study of channels that have not been investigated in this context, or 
that have not been considered promising because of the expected large jet-to-$\tau$ background.  
Assuming the experiments can achieve similar tau reconstruction performance in $\sqrt{s} = 14$ TeV 
data as they have in $\sqrt{s} = 7$ TeV data, each experiment can measure the cross sections of 
$WH$ and $t\bar{t}H$ production in the $H\rightarrow \tau\tau$ decay channels to $\approx 20\%$ 
precision with 100 fb$^{-1}$ of integrated luminosity.  Additionally, with achievable $\met$ 
reconstruction, a measurement of $ZH$ production with an accuracy of $\approx 25\%$ is possible 
with the same luminosity.  With more data the sensitivity to $ZH$ and $t\bar{t}H$ production 
should improve, while sensitivity to $WH$ production is unlikely to improve significantly due to 
systematic uncertainties on the background.  If the assumed $\tau$ identification efficiency or 
$\met$ resolution cannot be achieved, targeted background rejection through e.g. a multivariate 
analysis or improved mass reconstruction could compensate.  Including additional decays of the 
tau leptons or top quarks would also improve sensitivity.  By combining the associated production 
measurements in $H\rightarrow \tau\tau$ decays with measurements of the same production mechanisms 
in $H\rightarrow b\bar{b}$ decays~\cite{tthbb,jetstructure, multivariate}, a precision of 
$\approx 20\%$ on the ratio of partial widths $\Gamma_{\tau}/\Gamma_b$ is achievable.  We expect 
associated Higgs production with $H\rightarrow \tau\tau$ to provide an important contribution to 
Higgs coupling measurements with 100 fb$^{-1}$ of integrated luminosity at $\sqrt{s} = 14$ TeV.

This work was supported by the Science and Technology Facilities Council of the United Kingdom.  
CH would like to thank M. Mulhearn for discussions leading to the initial study of $WH$ production.

\end{document}